\documentstyle[preprint,eqsecnum,aps,epsf]{revtex} 
\tightenlines

\newcommand{ \bq }{{\bf q}} 
\newcommand{ \br }{{\bf r}}

\begin{document} 
\draft 
\preprint{LBNL-43486} 
 
\title{The relationship between  particle freeze-out distributions  
and HBT radius parameters} 
\author{D. Hardtke$^{(a)}$ and S.A. Voloshin$^{(a,b)}$} 
\address{ 
(a) Nuclear Science Division, Lawrence Berkeley National Laboratory, 
Berkeley, CA  94720 
\\ 
(b) Department of Physics and Astronomy, Wayne State University, 
Detroit, MI 48201 
}

\date{\today} 
\maketitle\begin{abstract} 
The relationship between pion and kaon space-time freeze-out distributions 
and the HBT radius  parameters in high-energy nucleus-nucleus 
collisions is investigated.   
We show that the HBT radius parameters in general do not reflect  
the R.M.S. deviations of the single particle production points.   
Instead, the HBT radius parameters are most closely related  
to the curvature of the two-particle space-time  
relative position distribution at the origin.    
We support our arguments by 
studies with a dynamical model (RQMD 2.4).  
\end{abstract} 
\pacs{} 
 
\narrowtext 
\section{Introduction} 
\label{intro} 
 
Two-particle intensity interferometry (HBT) has  
been used extensively in nucleus-nucleus  
collisions to measure the space-time extent of the particle emitting  
region\cite{HeinzJacak}.  
Because of the complicated dynamics of high-energy nucleus-nucleus collisions, 
the HBT radii, defined as parameters of a Gaussian fit to the 
correlation function, 
do not measure the full size of the particle emitting region  
directly. Instead, the HBT radius parameters reflect an interplay  
between the geometric system size, the 
expansion of the system, the duration of particle emission,  
and the contribution of particles from resonance decays.  It is hoped 
that the two-particle correlation functions can be parameterized in such a way 
that the various factors that influence the measured HBT radii can be  
understood individually. 
 
The two-particle correlation function is typically  
measured as a function of momentum 
difference $q=p_1-p_2$ and parameterized in term of radius  
parameters $R_i$ and the so-called haoticity parameter $\lambda$:\footnote{ 
For simplicity, we omit in this parameterization the possible 
cross-terms~\cite{HeinzJacak}, 
which are not important for our study of particle production at {\em 
mid-rapidity} and in {\em central} collisions.}  
\begin{equation} 
C_2(q) = 1+\lambda e^ 
{-q_i^2R_i^2}. 
\label{ec2} 
\end{equation} 
For a static Gaussian pion source, the HBT radius parameters $R_i$  
directly correspond to the variance of the particle spatial  
distribution at freeze-out~\cite{HeinzJacak},  
\begin{equation} 
R_i^2 = RMS^2_{(x_i-V_it)} = \sigma^2(x_i-V_it), 
\label{statGauss} 
\end{equation} 
where $V_i$ is the particle velocity and $R_i$ are the HBT radius parameters 
obtained from a fit to the correlation function.   
For a non-Gaussian source, this simple relationship no longer holds.   
In this paper, we explore the meaning of the Gaussian radius 
parameters in the case of a non-Gaussian particle source 
and use a dynamical model to investigate the breakdown  
of equation \ref{statGauss}.

\section{Basic relations} 
 
Note that in general case of the particle source function the relation 
(\ref{statGauss}) {\em cannot} be true.  
A simple example would be the source with added a few particles 
emitted at very large distances. 
The addition of such particles can significantly change  
the freeze-out distribution R.M.S value 
and at the same time should not change the 
correlation function (sensitive only to pairs with close 
production points). 
Below we argue that the correlation function, and in particular the 
HBT radii derived from the Gaussian fit to the correlation function  
are indeed sensitive only to the distribution of close pairs.  
Hence, they can be quite different from the R.M.S. values of particle 
freeze-out distribution.  
 
Let us show that  
the Gaussian fit to the correlation function yields the value of  
$\langle q_i^2  \rangle = (2R_i^2)^{-1}$ where the average is taken 
 over the function  $C_2(q)-1$.  
Consider a Gaussian fit to a one-dimensional correlation function 
using the maximum likelihood method.  The normalized fit function is 
\begin{equation} 
f(q) =  \frac{2}{\sqrt{\pi}}Re^{-q^2R^2}.  
\end{equation} 
The likelihood functional to be minimized is then 
\begin{equation} 
\ln L = \ln \prod_i f(q_i) =  
\ln \prod_i (\frac{2}{\sqrt{\pi}}Re^{-q_i^2R^2}) =  
\sum_i (\ln\frac{2}{\sqrt{\pi}} + \ln R - q_i^2R^2)  
\end{equation} 
The likelihood equation becomes, 
\begin{equation} 
\frac{\partial \ln L}{\partial (R^2)} = \sum_i (\frac{1}{2 R^2} - q_i^2) = 0. 
\end{equation} 
Solving this equation yields, 
\begin{equation} 
\frac{1}{2R^2} = \frac{\sum_i q_i^2}{N} = \langle q^2 \rangle. 
\label{eqn:fitR} 
\end{equation} 
Hence, irrespective of the shape of the measured correlation function, the  
fitted Gaussian HBT radii measure the value of $\langle q^2 \rangle$. 
 
$\langle q^2 \rangle$ can also be related to the shape of the source 
function (time integrated relative distance distribution) 
$S(\Delta \bf{r})$, where $\Delta \bf{r}$ is the distance between 
particle production points $r_i = x_i - V_it$.   
Using this function, the  
correlation function in momentum space 
can be Fourier transformed into position space\cite{Brown,Wiedemann}:   
\begin{eqnarray} 
C(q) - 1 & = & \int e^{-i\bq \Delta \br } S(\Delta \br) d\Delta \br \\ 
S(\Delta \br) & = &  
\frac{1}{(2\pi)^3}  
\int e^{i\bq \Delta \br} (C(q)-1) d \bq. 
\label{eqn:S} 
\end{eqnarray}  
Differentiating the source function twice yields, 
\begin{equation} 
\frac{1}{S(\Delta \br)} 
\left. \frac{\partial^2 S(\Delta \br)}{\partial  
\Delta r_i^2} \right|_{\Delta \br = 0} =  
- \langle q_i^2 \rangle, 
\label{eqn:deltaS} 
\end{equation} 
where the average is taken over the correlation function, 
$C(q)-1$. 
Eqns. \ref{eqn:fitR} and \ref{eqn:deltaS} show that the HBT radius  
parameters are directly related 
to the second derivative of the source function of relative  
production points near $\Delta \br = 0$.   
This is true irregardless of the shape of the correlation function.   
 
Note that the lambda parameter in the fit function (Eq. \ref{ec2}) 
in a maximum likelihood approach is defined by the normalization 
\begin{equation} 
\int d\bq \lambda e^{-q_i^2 R_i^2} =\int (C(q)-1) d\bq. 
\end{equation} 
The r.h.s. of this expression can be determined from equation 
\ref{eqn:S}: 
\begin{equation} 
\int (C(q)-1) d\bq = (2\pi)^3 S(0). 
\end{equation} 
It follows then that 
\begin{equation} 
\lambda= 2^3 \pi^{3/2} S(0) R_1 R_2 R_3, 
\end{equation} 
where $S(0)$ is the (true) source function at the origin, and $R_1$, 
$R_2$, and $R_3$ are the (fitted) radius parameters. 
 
\section{Model Calculations} 
 
To understand the relationship between the fitted HBT radii and the source 
function of relative production points and check how well the above equations 
work for a realistic particle source, 
the event generator RQMD (v2.4) \cite{RQMD} is employed.  This  
model has been shown to reproduce experimental HBT data reasonably well 
\cite{sullivan}.   
We focus on  particle production in midrapidity region $(|y|<0.5$) in 
central ($b<3$fm) Au+Au collisions at RHIC 
energy ($\sqrt{s} = 200$~GeV/nucleon).   
 
\begin{figure}[htbp] 
\begin{center} 
\leavevmode 
\epsfysize=13cm 
\epsfbox{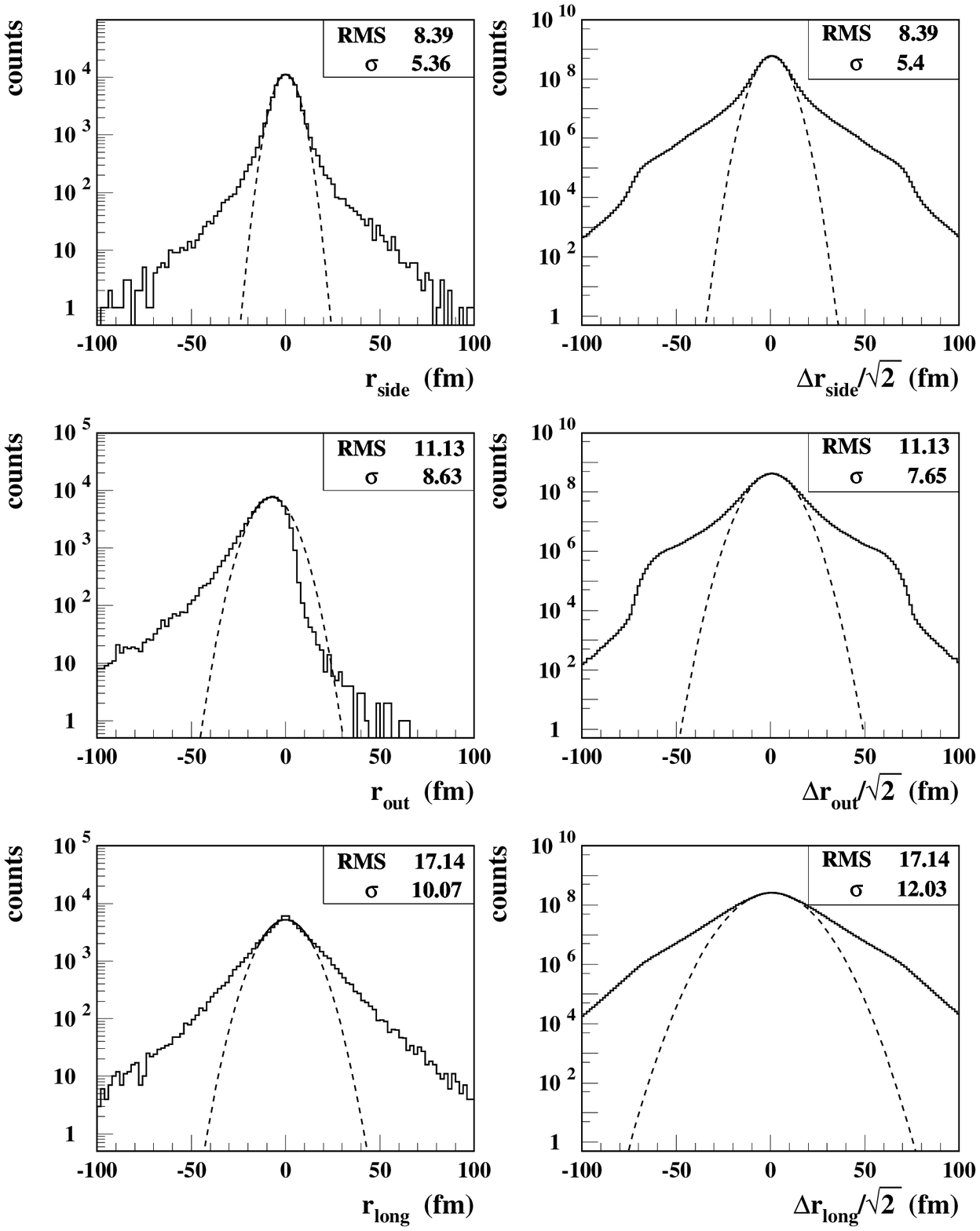} 
\end{center} 
\caption{The single particle freeze-out position distributions and 
two-particle  position difference distributions for pions 
with $0<p_T<200$MeV/c and $|y|<0.5$ for  $r_{side}$ 
(perpendicular to the transverse momentum vector), $r_{out}$  
(parallel to the transverse momentum vector), and $r_{long}$ (along the beam 
axis).  The Gaussian fit function is also shown in each case, along with the 
fit parameters.} 
\label{Freezeout} 
\end{figure}

Figure \ref{Freezeout} shows single  
particle freeze-out position distribution  
and the two-particle relative position  
distribution for pions.  
The single particle freeze-out position difference distribution 
has long non-Gaussian tails.  In addition, the shape is not well described 
by a Gaussian at $r=0$.  The two-particle position difference distribution  
still shows large non-Gaussian tails.  The distribution, however, is well 
described as a Gaussian near $\Delta \bf{r} = 0$.  Fitting this 
distribution to  
a Gaussian (also shown in the Figure) yields a Gaussian width $\sigma$ that is 
substantially smaller than the R.M.S. deviation of the distribution. 
 
The two-particle correlation functions 
are calculated as a function of $p_T$ for mid-rapidity ($|y|<0.5$)  
pions and kaons assuming plane wave propagation for the outgoing particles.   
The three-dimensional correlation function is then fit using the Bertsch-Pratt 
out-side-long parameterization, 
\begin{equation} 
C_2(q_{o},q_{s},q_{l}) = 1+\lambda e^{-q_o^2R_o^2-q_s^2R_s^2-q_l^2R_l^2},  
\end{equation} 
where $q_l$ is the momentum difference along the beam axis, $q_o$ is the  
momentum difference perpendicular to the beam axis and parallel to the total 
transverse momentum of the pair, and $q_{s}$ is the momentum difference  
perpendicular to the beam axis and perpendicular to the total transverse  
momentum of the pair.   
 
\begin{figure}[htbp] 
\begin{center} 
\leavevmode 
\epsfysize=12cm 
\epsfbox{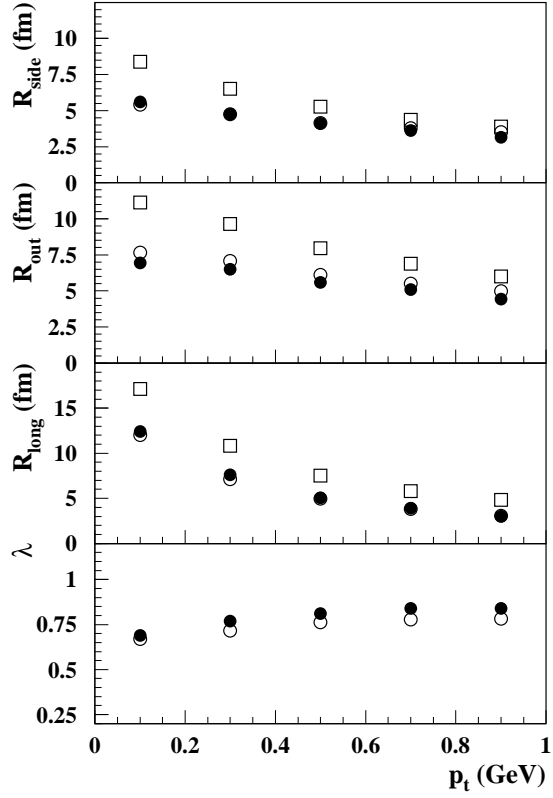} 
\end{center} 
\caption{The fitted HBT radius parameters (solid circles), 
single-particle 
R.M.S. deviations (open squares), and the sigma of the Gaussian 
fit two-particle position difference distribution (open circles) 
as a function of $p_t$ for  $R_{side}$, $R_{out}$, $R_{long}$.
The rapidity range is $|y|<0.5$.
The bottom plot show the lambda parameter determined from the fit to
the correlation function (solid circles) and from the two particle
position difference distribution.}
\label{Rvspt} 
\end{figure}

Figure \ref{Rvspt} shows the fitted HBT  
radius parameters as a function of transverse momentum.  Also plotted are the 
R.M.S. deviations of the single particle freeze-out distributions and the  
fitted Gaussian widths 
of the two-particle position difference distributions.  
The R.M.S. deviations are substantially larger than the fitted HBT radii 
in all cases, while the fitted Gaussian widths of the two-particle relative  
position difference distributions are very close to the fitted HBT radii. 
The single-particle R.M.S. deviation differs from the fitted HBT radii most  
at low transverse momentum.  This deviation arises from the long non-Gaussian 
tails in the freeze-out position distribution due to pions from resonance  
decay and particular expansion dynamics of the system.   
 
The width parameter (sigma) of the Gaussian fit to 
the two-particle difference distribution 
near the origin (in our calculations we perform the fit in the region 
of $\pm 0.5$~R.M.S.) 
is a very good estimate of the second derivative of the relative  
source function at $\Delta \bf{r}=0$.   
Hence, the observed good agreement between HBT radii (fit parameters 
to the correlation function) and widths of the Gaussians fitted to the 
 two-particle position difference distribution near the origin 
confirm  relationship shown in eqns. \ref{eqn:fitR} and \ref{eqn:deltaS}. 
Note a simple interpretation of the lambda parameter following from 
Fig.~1 (two-particle position difference distributions). 
It is just a product of the ratios of area under the Gaussian fit and 
the total number of pairs used for the distribution. 
 
\begin{figure}[htbp] 
\begin{center} 
\leavevmode 
\epsfysize=11cm 
\epsfbox{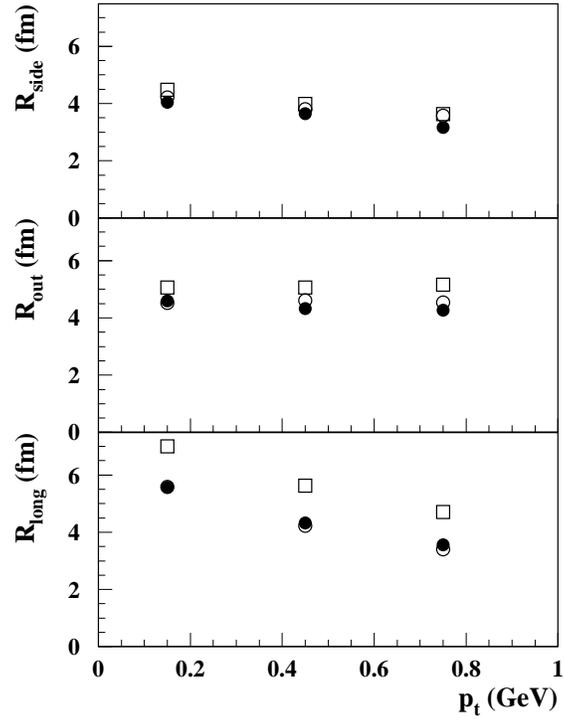} 
\end{center} 
\caption{The same as Fig. \protect\ref{Rvspt}, but for kaons.} 
\label{Rvsptk} 
\end{figure}

Figure \ref{Rvsptk} is similar to figure \ref{Rvspt} except that the  
particles used are kaons.  In this case, the R.M.S. deviations of the  
single-particle freeze-out distributions and the fitted Gaussian width of the  
two-particle relative distance distribution are very similar in the transverse 
directions.  Since fewer kaons come from resonance decays, the long  
non-Gaussian tails in the single particle freeze-out distributions are  
greatly reduced.   
In the longitudinal direction, where the non-Gaussian nature of the 
production point distribution is mostly defined by the system 
expansion dynamics (very similar for pions and kaons), the difference 
between the HBT radii and the R.M.S. values of freeze-out 
distribution is quite significant.    
The lambda parameter in the kaon case is generally larger 
than that for the pion 
case but due to a limited statistics exhibits large fluctuations and 
is omitted from the figure.

\section{Conclusions} 
 
We have shown that the HBT radius parameters obtained from a Gaussian fit 
to the correlation function are equivalent to the  
the curvature of the relative  
freeze-out distributions at $\Delta \bf{r} = 0$.   
Therefore, 
the fitted HBT radii are directly related to  
fitted Gaussian width of the two-particle position difference  
distribution near the origin.   
The lambda parameter is determined by the ratio of the value of the 
real relative distance distribution at the origin and the distribution 
chosen in a Gaussian form with the radii parameters determined by the 
fit to the shape of the correlation function. 
Studies with a dynamical model support the above statements.

\acknowledgments 
 
We would like to thank Heinz Sorge for the use of his RQMD code, and Scott  
Pratt for the use of the CRAB correlation after-burner.  In addition we 
would like to thank Nu Xu for providing the RQMD 
events and useful discussions and U.~Heinz for comments.   
 
This work was supported by the Director, Office of Energy Research, 
Office of High Energy and Nuclear Physics, Division of Nuclear Physics 
of the U.S. Department of Energy under Contracts DE-AC03-76SF00098 and 
DE-FG02-92ER40713.

\end{document}